%% file: main.tex
\documentclass{article}

\usepackage{arxiv}
\usepackage{cite}
\usepackage{graphicx} 
\graphicspath{ {images/} }
\usepackage{subcaption}
\usepackage{amsmath}
\usepackage[capitalise]{cleveref}
\usepackage{url}
\usepackage[symbol, multiple]{footmisc}

\usepackage{tikz}
\newcommand\copyrighttext{%
  \footnotesize 
  This is the version of the article after acceptance and before editing,
  as submitted by an author to \textit{Neuromorphic Computing and Engineering}.
  IOP Publishing Ltd is not responsible for any errors or omissions in this version of the manuscript or
  any version derived from it. The Version of Record is available online at
  \url{https://doi.org/10.1088/2634-4386/acfe36}.
  }
\newcommand\copyrightnotice{%
\begin{tikzpicture}[remember picture,overlay]
\node[anchor=south,yshift=10pt] at (current page.south) {\fbox{\parbox{\dimexpr\textwidth-\fboxsep-\fboxrule\relax}{\copyrighttext}}};
\end{tikzpicture}%
}

\usepackage{verbatim}

\usepackage{siunitx}

\usepackage{ulem}

\renewcommand{\headeright}{}
\renewcommand{\undertitle}{}
\renewcommand{\shorttitle}{Time-coded spiking FT}

\begin{document}

\title{Integrate-and-fire circuit for converting analog signals
to spikes using phase encoding}

\date{}
\author{Javier~López-Randulfe\thanks{Corresponding author. E-mail: lopez.randulfe@tum.de} \thanks{
TUM School of Computation, Information, and Technology. Technical University of Munich, Munich, Germany} ,
        Nico Reeb\footnotemark[2] ,
        \textbf{and~Alois~Knoll\footnotemark[2]}
}

\maketitle

\input{00-abstract}
\copyrightnotice
\input{01-introduction}

\input{02-method}
\input{03-Results}
\input{04-discusion}
\input{05-conclussion}

\input{XX-outro}

\end{document}

%% file: 00-abstract.tex
\begin{abstract}

Processing sensor data with spiking neural networks
on digital neuromorphic chips
requires converting continuous analog signals into 
spike pulses.
Two strategies  are promising for achieving low energy consumption and fast processing speeds 
in end-to-end neuromorphic applications.
First, to directly encode analog signals to spikes to bypass
the need for an analog-to-digital converter (ADC).
Second, to use temporal encoding techniques to maximize the spike sparsity,
which is a crucial parameter for fast and efficient neuromorphic processing.
In this work, we propose an adaptive control of the refractory period of the leaky integrate-and-fire (LIF) neuron model
for encoding continuous analog signals into a train of \mbox{time-coded} spikes.
The LIF-based encoder generates phase-encoded spikes that are compatible with digital hardware.
We implemented the neuron model on a physical circuit
and tested it with different electric signals.
A digital neuromorphic chip processed the generated spike trains
and computed the signal's frequency spectrum using a spiking version of the Fourier transform.
We tested the prototype circuit on electric signals up to 1 KHz.
Thus, we provide an end-to-end neuromorphic application that generates the
frequency spectrum of an electric signal without the need for an ADC or a digital signal processing algorithm.

\end{abstract}


%% file: 01-introduction.tex
\section{Introduction}
\label{sec:introduction}

The human brain processes a vast amount of sensory information
quickly and efficiently. 
This efficient and fast processing is currently unmatched by any artificial system.
Hence, ongoing research aims to understand and mimic the brain's sensory processing.
An essential part of this research is encoding incoming information 
into spikes \cite{brette2015philosophy}.
Coding schemes are commonly classified into rate coding and temporal coding, that assume 
information is stored in the rate of spike trains for a given time window or
in the actual timing of the spikes, respectively.  
There is evidence that points toward the existence of time coding schemes in the brain \cite{bair1996, reich2000, brasselet2012}, as well as for rate coding schemes \cite{Hubel1968ReceptiveFA, heeger_1992, london2010}.
These two paradigms can also be generalized for neuron populations, where redundant information in a group of neurons allows to reduce the inference time or the uncertainty in the signal \cite{pouget2003}.
From a conceptual point of view, rate encoding offers robustness and can be easily generalized for large populations \cite{london2010}.
In contrast, time coding is faster and more energy efficient, as it requires few spikes.
It also explains the fast reaction times, especially for sensory pathways, that are crucial for survival
\cite{heil2004, gollisch2009, martin_davis_riesenhuber_thorpe_2018}.
One hypothesis of how neurons implement time coding is by carrying information in the inter-spike interval,
i.e., the time between two consecutive spikes \cite{oswald2007interval}.
Alternatively, time coding schemes may rely on reference points in time
to convey information. These references include the onset of a stimulus,
spikes from other neurons, or periodic waves within the brain activity \cite{buzsaki2013, hakim2018}.

One of the aims of neuromorphic research is to represent 
data with discrete spikes, mimicking the efficient operation of the brain.
Therefore, spike encoding is a crucial step in the design of neuromorphic algorithms.
When developing spiking neural networks (SNNs), 
the choice of the encoding scheme depends on aspects like the nature of the input data,
the network architecture, or the neuron dynamics model used by the SNN.
Neuromorphic computing research often focuses on high-end tasks that use processed data,
e.g., classification \cite{rueckauer2017conversion}, object detection \cite{lopez2021spiking}, tracking \cite{cao2015spiking}, or motor control \cite{cao2015spiking, stagsted2020towards}.
These applications typically use rate encoding,
as the main motivation is to validate neural models in terms of accuracy,
rather than the benefits of sparsity and efficiency provided by temporal encoding schemes \cite{davidson2021comparison}.
However, energy and time efficiency becomes more relevant for low-level neuromorphic applications
that directly process sensor data.
This is the case of embedded systems where the pool of energy is limited,
such as automotive applications\cite{zhou2020deep, vogginger2022automotive}.
There are recent examples of neuromorphic computing algorithms that deal with low-end tasks
and are applied to sensor data,
such as LiDAR \cite{zhou2020deep}, event-based cameras \cite{ranccon2022stereospike}, FMCW radar \cite{vogginger2022automotive},
electrocardiogram signals \cite{gerber2022neuromorphic}, or microphones \cite{auge2021end}.

The interface between the sensor and the neuromorphic chip is crucial for obtaining a high energy efficiency and processing speed.
When dealing with analog data, the most efficient approaches directly operate on the sensor signals
by using ad-hoc circuits that encode them to spike trains.
The signal-to-spike encoding mechanisms are typically application-specific, as they need to be compatible with the SNNs afterward.
The idea of encoding analog signals to spikes for obtaining efficient implementations is not new.
In \cite{gerber2022neuromorphic} and \cite{burelo2022neuromorphic}, authors use temporal contrast to identify positive and negative gradients,
which discriminate high and low frequencies in the analog signal.
\cite{zhao2017interspike} introduces a VLSI circuit for implementing the interspike-interval (ISI) mechanism for encoding analog signals.
The authors in \cite{livi2009current} and \cite{wijekoon2008compact} introduce silicon neurons for generating different spike bursting behaviours,
similar to those observed in biological neurons.
We refer the interested reader to \cite{chicca2014neuromorphic} for getting an overview of the different analog circuits used in neuromorphic systems for real-time applications,
and to \cite{forno2022spike} for a comparison of the different encoding approaches.

Some neuromorphic applications focus on generating the frequency spectrum of the incoming signal \cite{auge2021end, orchard2021efficient, lopez2022time}.
In these examples, an \mbox{analog-to-digital} converter (ADC) samples the sensor data and an SNN computes a higher-level algorithm
on a digital neuromorphic chip afterward.
To maximize the efficiency of these applications, 
the ADC should be removed by directly encoding the input signal into a spike train compatible with the digital neuromorphic chip.
For the specific case of an SNN that implements the Fourier transform (FT) \cite{lopez2022time},
the network needs spikes referenced to a periodic signal,
as the FT algorithm works with data sampled uniformly over time.
Thus, the best-suited temporal approach for these applications is phase encoding,
as ISI and temporal contrast lack a signal that serves as reference over the time dimension.
Additionally, the results in \cite{forno2022spike} show the benefits of
phase encoding in terms of accuracy and efficiency.
The benefits are more clear when comparing
with rate-coded approaches,
as phase coding needs down to $6.5$ times less 
spike operations for processing data \cite{guo2021neural},
resulting in a lower energy footprint
\cite{davidson2021comparison}.

The work in \cite{lopez2022time} proposed an SNN that computes a mathematically equivalent version of the FT by using time encoding.
The authors implemented the spiking FT (S-FT) on a digital neuromorphic chip and validated it on automotive radar data.
Here, we propose adapting the leaky integrate-and-fire (LIF) neuron model for directly encoding analog signals
to  phase-encoded spikes, which the S-FT can use on a digital neuromorphic chip.
The proposed approach leads to the implementation of an end-to-end pipeline
that does not require an ADC
nor a digital processing algorithm that encodes digitized signals into spikes.
The proposed \mbox{analog-to-spike} encoder (ASE) employs  phase encoding,
i.e., each sample is represented by the time difference between the spike and a reference pulse.
We implemented the ASE on a physical circuit
and validated the system with real data by comparing the reconstructed data from the spikes with the original data.
We modeled the nature of the error of the ASE for different scenarios.
Moreover, we implemented the S-FT on the neuromorphic chip \mbox{SpiNNaker 2} \cite{hoppner2021spinnaker} and ran it with the ASE output spikes.
The results show that the chip can use the spike train that results from encoding a periodic wave.
To our knowledge, we provide the first \mbox{end-to-end} neuromorphic pipeline
for generating the frequency spectrum of analog signals without employing an ADC.

%% file: 02-method.tex
\section{Analog-to-spike encoder}
\label{sec:ase}

The LIF is a century-old single-compartment neuron model that mimics the behavior of biological neurons \cite{abbott1999lapicque}.
Its simplicity makes it a widely used artificial neuron in neuromorphic engineering.
The model is based on an $R-C$ circuit that charges with the incoming current and discharges in the absence of input,
resembling the flux of ions in a biological neuron through its dendrites and the constant leak of them over time through its membrane.
We can express the dynamics of the LIF as
\begin{equation}
    \label{eq:lif_current}
    C \frac{du}{dt} + \frac{u(t) - u_\text{rest}}{R} = i_\text{in}(t) \,,
\end{equation}
where $u(t)$ is the neuron's membrane potential, $u_\text{rest}$ is the
resting potential, and $i_\text{in}(t)$ is the input current to the neuron.
The LIF model does not include the depolarization and hyperpolarization mechanisms  on a biological neuron when it generates a spike.
Instead, it is necessary to manually add the behavior of the spike generation and refractory period.
A spike is triggered at time $t_s$ when $u(t)$
crosses the threshold voltage
$u_\text{th}$.
After the spike generation, $u(t)$ is reset and fixed to
$u_\text{rest}$ for the neuron's refractory period $t_\text{refr}$,
\begin{align}
    \label{eq:lif_spike}
    &t_s : u(t_s) \geq u_\text{th} \quad \text{and} \\
    &u(t)=u_\text{rest} \quad \text{for} \quad t_s < t < t_s + t_\text{refr} \,.
\end{align}

As artificial sensors often provide information in the form of a voltage,
we modify \eqref{eq:lif_current} so the evolution of the membrane voltage depends on an input voltage $u_\text{in}$.
We also set $u_\text{rest} = 0$ for simplicity,
\begin{equation}
    \label{eq:lif_voltage}
    RC \frac{du}{dt} + u(t) = u_\text{in}(t) \,.
\end{equation}
In the remainder of the paper, we use the time constant \mbox{$\tau=RC$}.
By integrating \eqref{eq:lif_voltage} and setting $u(t_s) = u_\text{th}$,
we obtain an encoding function $f(u_\text{in})$ that maps
the input voltage $u_\text{in}$ to a spike time $t_s$ after onset,
\begin{equation}
    \label{eq:spike_time}
    t_s = f(u_\text{in}) = - \tau log\left(1-\frac{u_\text{th}}{u_\text{in}}\right) \,.
\end{equation}
Accordingly, the ideal decoding function is given by the inverse of $f(u_\text{in})$,
\begin{equation}
    \label{eq:ideal_decoding}
    u_\text{in} = f^{-1}(t_s) = \frac{u_\text{th}}{1-e^{-t_s/\tau}}  \,.
\end{equation}

We implemented the LIF neuron model \eqref{eq:lif_voltage} for generating a spike train from input analog signals.
We encoded the information into precise spike times using \eqref{eq:spike_time} and controlling the refractory period according to a reference signal.

\begin{figure*}[h]
    \centering
    \includegraphics[width=\textwidth]{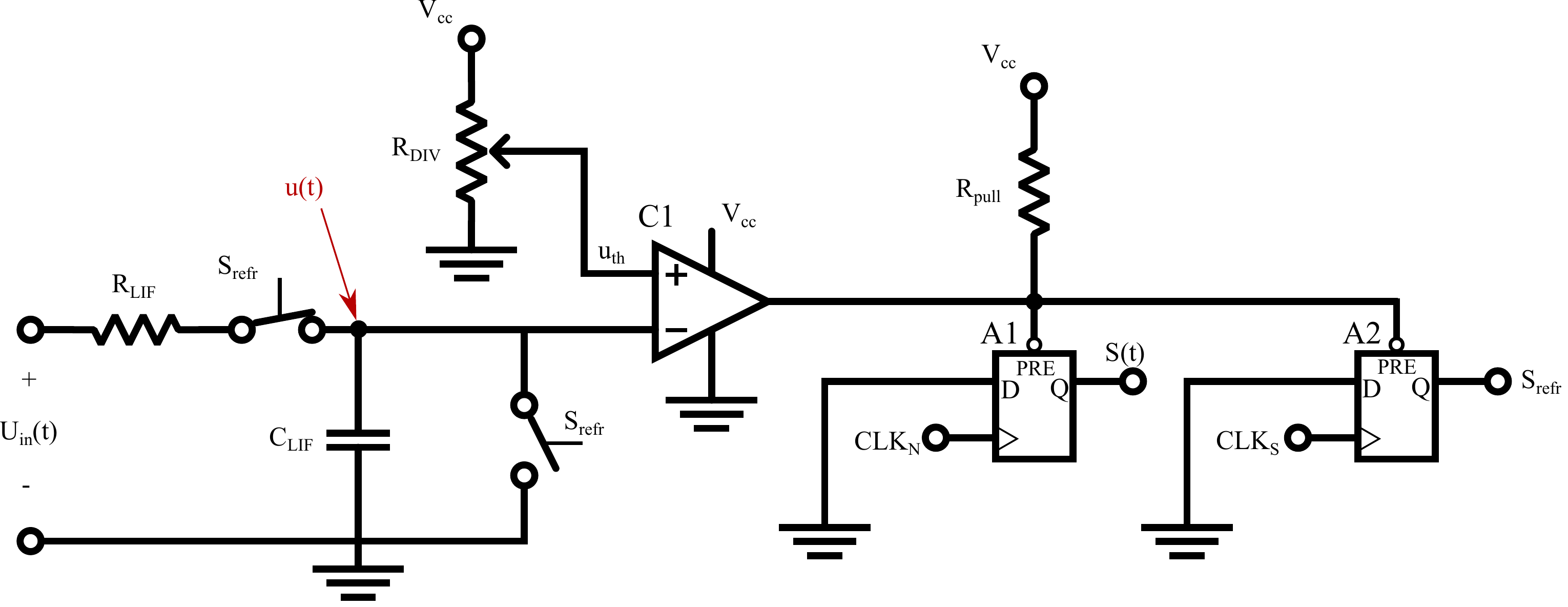}
    \caption{Simplified schematic of the electric circuit used for implementing the LIF-based encoder.
    The $R-C$ formed by $R_\text{LIF}$ and $C_\text{LIF}$ circuit generates the $u(t)$ according to \eqref{eq:lif_voltage}.
    $C1$ compares $u(t)$ with $u_{th}$ and generates spike events.
    The two D-type flip-flops synchronize the output of $C1$ with the spike reading hardware,
    which controls the reset of the spike s(t) and refractory $S_\text{refr}$ signals.
    The second flip-flop regulates the working cycle of the LIF circuit.}
    \label{fig:electric_schematic}
\end{figure*}

\cref{fig:electric_schematic} shows a simplified schematic of the electric circuit we used to implement the encoder.
This circuit served as a proof-of-concept for implementing an end-to-end pipeline.
For implementing a high-performance circuit, we refer the interested reader
to literature focused on the VLSI design of silicon neurons \cite{chicca2014neuromorphic}.

\subsection{Encoding analog voltages with temporal spikes}

By default, the LIF dynamics described in \eqref{eq:lif_voltage} generates a spike rate proportional to the input voltage $u_\text{in}(t)$.
In our approach, we sample the continuous signal $u_\text{in}(t)$
with a constant sampling time  $T_S$
by generating a single spike per sample.
We repeat the spike generation process for successive time windows.
Thus, the $m^\text{th}$ spike represents the signal $u_\text{in}$
during the time range \mbox{$[t_m, t_m+T_S]$}.
This is inspired by the phase encoding technique hypothesized for the encoding of information in certain areas of the brain \cite{hakim2018}.
A binary variable $S_\text{refr}$ forces $u(t)$ to stay in a refractory state,
\begin{equation}
    \label{eq:refractory}
    u(t) = u_\text{rest}, \quad \mbox{iff} \; S_\text{refr}=1 \,.
\end{equation}
We achieve an adaptive refractory period, $t_\text{refr} = T_S - t_s$,
by controlling 
the set and reset time of $S_\text{refr}$ for the $m^\text{th}$ sample,
\begin{equation}
  S_\text{refr} =
    \begin{cases}
      1 & \text{if } t_m+t_s < t < t_m+T_S\\
      0 & \text{otherwise.}
    \end{cases}
\end{equation}
This encoding mechanism requires a periodic 
reference signal $CLK_S$ (see \cref{fig:electric_schematic})
that defines the sampling time $T_S$ and the end of the refractory period.
We assume the input voltage to be constant during the sampling time of the ASE,
\mbox{$u_{in}(t) = U_{in}$}, as the neuron dynamics is faster than the rate of change of the input signal.

\begin{figure}[h]
    \centering
    \includegraphics[width=0.6\columnwidth]{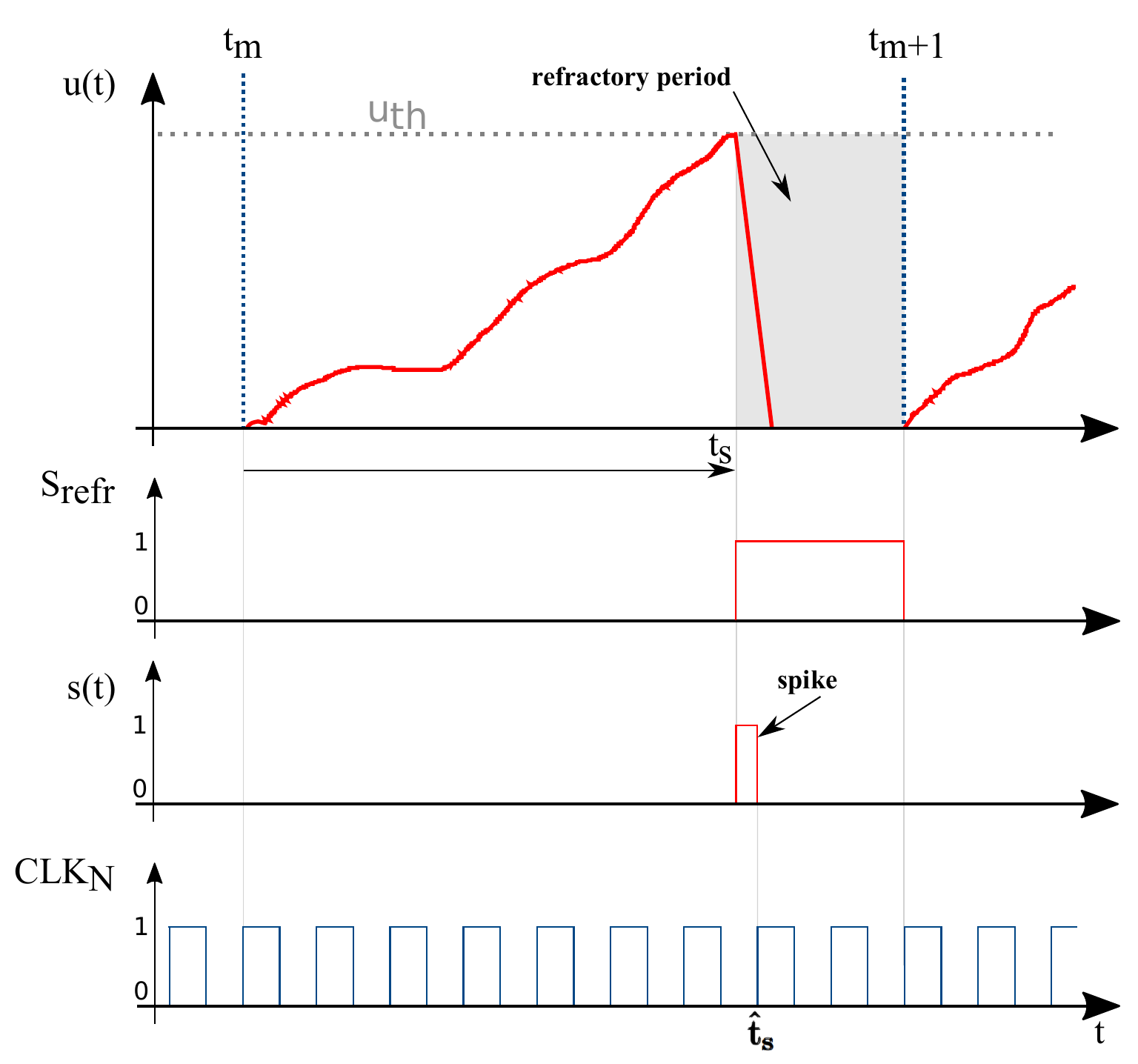}
    \caption{ASE signals for the generation of one spike.
    The top plot represents the membrane voltage for the n-th spike in the time series,
    the second plot represents the refractory state of the neuron,
    the third plot represents the output spike train,
    and the bottom plot represents the sampling clock of the digital neuromorphic chip that uses the spikes generated by the ASE.}
    \label{fig:ASE_signals}
\end{figure}

The sampling time  $T_N$ of the hardware component collecting the spikes determines the spike time resolution $N$,
\begin{equation}
    \label{eq:spike_resolution}
    N = \frac{T_S}{T_N} \,.
\end{equation}

\subsection{Parameter tuning}
\label{sec:tuning}

\begin{figure}
    \centering
    \includegraphics[width=0.6\textwidth]{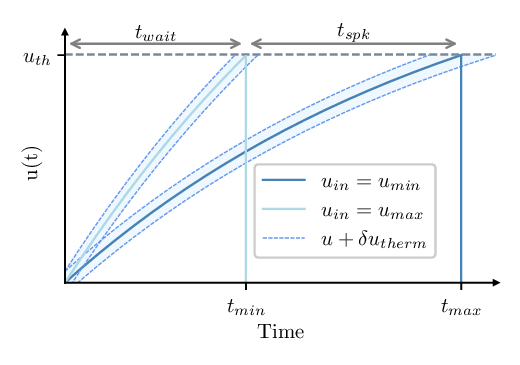}
    \caption{
    Voltage dynamics of the ASE for the corner cases.
    The plot shows the curves for the maximum and minimum possible inputs,
    which generate spikes at the minimum and maximum spike times, respectively.
    The dashed lines represent the additive thermal noise \eqref{eq:thermal_noise}.}
    \label{fig:spike_curves}
\end{figure}

The encoder (\ref{eq:spike_time}) has two parameters
that we can tune in order to map 
a voltage range $\left[ u_{\text{min}}, u_{\text{max}} \right]$ 
onto a time range $\left[ t_{\text{min}}, t_{\text{max}} \right]$,
namely the time constant $\tau$, and the threshold voltage $u_\text{th}$.
Due to the dynamics of the neuron model, 
we need to choose $u_\text{min} > u_\text{th} > 0$
to ensure a finite time range. This leads to $t_\text{min} > 0$ 
(see \eqref{eq:spike_time}) and hence
introduces a waiting time before the first informative spike can arrive.
We can use $\tau$ to reduce $t_\text{min}$, but
this would also lead to a smaller time range, as 
$(t_\text{max} - t_\text{min}) \propto \tau$.
To optimize the resolution in the encoding time window $\Delta t$, 
we aim for a small waiting time 
\begin{align}
t_\text{wait} =  t_\text{min} = - \tau \ln \left( 1 - \frac{u_\text{th}}{u_\text{max}} \right)
\end{align}
and a large spike time range 
\begin{align}
    t_\text{spk} = t_\text{max} - t_\text{min} = - \tau \ln \left( 
    \frac{u_\text{max} (u_\text{min} - u_\text{th})}
    {u_\text{min} (u_\text{max} - u_\text{th})} \right).
\end{align}
Hence, by maximizing the ratio
\begin{align}
    \label{eq:time_ratio}
    \mu = \frac{t_\text{max} - t_\text{min}}{t_\text{min}} = \frac{t_\text{spk}}{t_\text{wait}} \, ,
\end{align}
we can ensure to find an optimal value for $u_\text{th}$ (see \cref{fig:spike_curves}).
The ratio \eqref{eq:time_ratio} is independent of $\tau$. Therefore,
we can only use $\tau$ as a scaling factor to fit the time range into the encoding window $\Delta t$.
A voltage threshold $u_\text{th} \sim u_\text{min}$ maximizes the ratio.

The overall encoding performance also depends
on the decoding and  the subsequent algorithms.
By tuning the encoding/decoding parameters, we optimize the process.
We establish three pipelines to evaluate the accuracy of the coding process.
First, a coding scheme using the proposed encoding $f(y)$
(\ref{eq:spike_time}) and its inverse as decoding $f^{-1}(t)$,
\begin{align}
\label{eq:inverse_encoding}
    y \xrightarrow{f} t \xrightarrow{f^{-1}} \hat{y} = y \,.
\end{align}
Second, a coding scheme using the proposed encoding $f(y)$
(\ref{eq:spike_time}) and a linear decoding $g^{-1}(t)$,
\begin{align}
\label{eq:approx_encoding}
    y \xrightarrow{f} t \xrightarrow{g^{-1}}  \hat{y} \,.
\end{align}
Finally, we assess the encoder performance
using a spiking algorithm $A_s$ that takes linearly encoded spike times as input
and produces linearly encoded spike times as output.
We apply the algorithm on spike times encoded as $t = f(x)$ \eqref{eq:spike_time}.
To evaluate the pipeline, we compare the decoded spiking output $g^{-1}(A_s(t)) = \hat{y}$
with the output of a non-spiking version of the algorithm $A(x) = y$,
\begin{align}
\label{eq:alg_encoding}
    & x \xrightarrow{f} t \mapsto A_s(t) \xrightarrow{g^{-1}}  \hat{y} \, .
\end{align}

For (\ref{eq:inverse_encoding}), the decoding does not need tuning because it
is the inverse of the encoding (\ref{eq:spike_time}). 
For (\ref{eq:approx_encoding}) and (\ref{eq:alg_encoding}), we
tune the parameters of the encoding and decoding to minimize the error 
\begin{align}
    \varepsilon = \int_{y_\text{min}}^{y_\text{max}} | \varepsilon(y)| dy = \int_{y_\text{min}}^{y_\text{max}} |y - \hat{y}| dy \, .
\end{align}

We evaluated the encoding on the S-FT algorithm \cite{lopez2022time}.
The S-FT network relies on linear coding and employs \mbox{current-based} LIF neurons without leak.
The model assumes that the input consists in phase-encoded spikes that follow
the encoding
\begin{align}
    \label{eq:ttfs_encoding}
    t_s &= g(y) = t_\text{lin, min} + \frac{t_\text{lin, max} - t_\text{lin, min}}{y_\text{max} - y_\text{min}} (y_\text{max} - y)
\end{align}
and decoding
\begin{align}
    \label{eq:ttfs_decoding}
    y &= g^{-1}(t_s) =  y_\text{max} - \frac{y_\text{max} - y_\text{min}}{t_\text{lin, max} - t_\text{lin, min}} (t_s - t_\text{lin, min}) \,.
\end{align}
By calculating the error 
\begin{align}
    \label{eq:linear_error}
    \varepsilon_\text{lin} = \int_{y_\text{min}}^{y_\text{max}} |y - \hat{y}| dy = \int_{y_\text{min}}^{y_\text{max}} |y - g^{-1}(f(y))| dy 
\end{align}
with the encoding $f(y)$ and decoding $g^{-1}(t)$,
we see that the multiplicative time constant $\tau$ cancels out.
The error depends on the decoder's time range $[t_\text{lin, min}, t_\text{lin, max}]$, the voltage threshold of the encoder $u_\text{th}$ and
the input voltage range $[u_\text{min}, u_\text{max}]$, where the latter is fixed by the given setup.
Due to the complexity of the relationship between the tunable parameters
$u_\text{th}$, $t_\text{lin, min}$ and $t_\text{lin, max}$, we propose a
loss function
\begin{align}
\label{eq:loss}
    \mathcal{L} = \alpha \varepsilon_\text{lin} -\mu \,,
\end{align}
that we can use to determine the optimal parameters, where $\alpha$ 
defines the weight of the linear error w.r.t. the time ratio, i.e. a small $\alpha$ 
puts focus on obtaining good time ratios, whereas big $\alpha$ favors 
small linear errors.
By setting a voltage threshold $u_\text{th}$ that
optimizes the time ratio (\ref{eq:time_ratio}), the decoding parameters
$t_\text{lin, min}$ and $t_\text{lin, max}$ can be determined by directly minimizing the
error $\varepsilon_\text{lin}$.

%% file: 03-Results.tex
\section{Validation experiments}
This section covers the experiment results for the proposed ASE.
We implemented a prototype on an electric board with standard components.
We tested the three scenarios introduced in section \ref{sec:tuning} with signals obtained from a function generator.
We collected the spikes and generated the reset signals for the encoder with a microcontroller from the dsPIC33CK family.
The minimum sampling time of the microcontroller is $1.48 \, \mu$s, which determines the resolution $T_N$ of the spike times.
We finally implemented the S-FT algorithm on the \mbox{SpiNNaker 2} chip \cite{hoppner2021spinnaker} and ran it with the spikes collected by
the microcontroller.

\begin{figure}
    \centering
    \includegraphics[width=0.6\columnwidth]{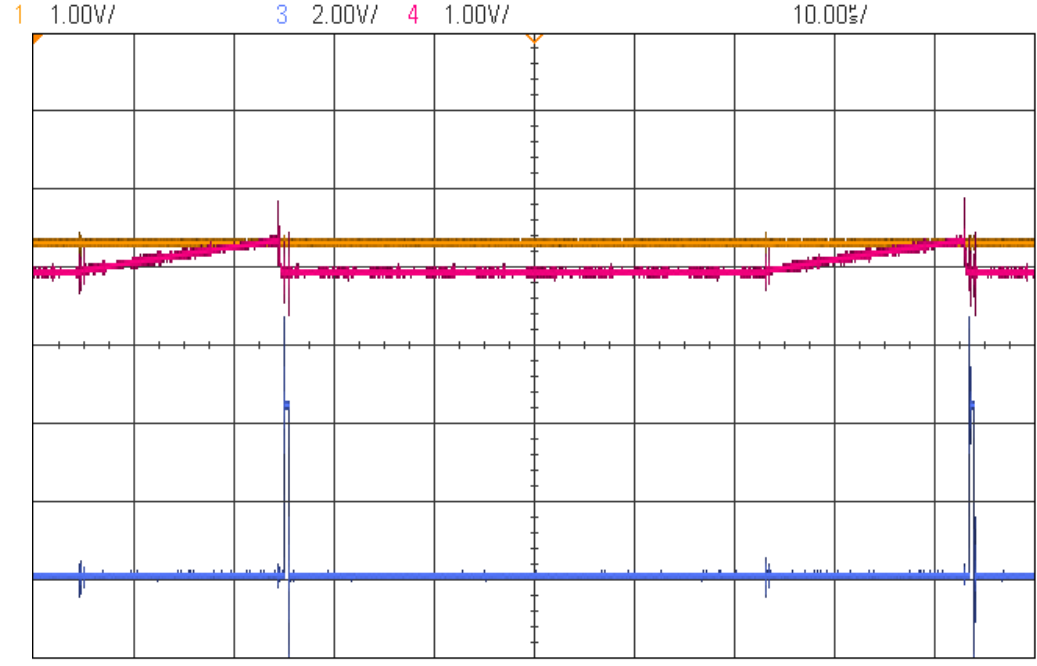}
    \caption{Measurement on a oscilloscope of the ASE membrane voltage $u(t)$,
    the threshold voltage $u_\text{th}$, and the spike event signal $s(t)$ in red, yellow, and blue, respectively.
    The measurements show the signals for two consecutive samples.}
    \label{fig:osilloscope_shot}
\end{figure}

We powered the circuit with $5$ V, and the input voltage signal was in the range of $[1, 5]$ V.
\cref{fig:osilloscope_shot} shows a measurement of the ASE on an oscilloscope for two consecutive spikes.

\subsection{Decoding of flat voltages with  an ideal decoder}
\label{sec:ideal_decoder}

We tested the ASE functionality using constant voltage signals as input.
We collected the generated spikes and applied the ideal decoding function (\ref{eq:ideal_decoding}) to the recorded spike times.
In this case, the absolute error in the measurement is mainly due to the quantization error and the circuit's thermal noise. 

The spike quantization introduces a time shift $\delta t_\text{q}$
to the actual spike time $t_s$ due to the finite resolution of the digital hardware reading the spike events.
Assuming that the spike time bins are uniformly distributed with a period of $T_N$,
we can theoretically model this error with an average sampling error
\begin{equation}
    \label{eq:quant_error}
    \delta t_\text{q} = \frac{T_N}{2} \,.
\end{equation}

On the other hand, voltage dynamics are noisy
and voltage fluctuations can alter the threshold crossing time.
The thermal error $ \delta t_\text{thermal}$ on the spike time highly depends on the voltage dynamics.
For simplification, we model the fluctuations as a constant added to the actual membrane voltage,
$u(t) + \delta u_\text{thermal}$, and we estimate its impact on the spike time as
\begin{equation}
    \label{eq:thermal_noise}
    t_s - \delta t_\text{thermal} = - \tau \ln \left( 1 - \frac{u_\text{th} - \delta u_\text{thermal}}{u_\text{in}} \right) \,.
\end{equation}
From \eqref{eq:thermal_noise}, we can conclude that the thermal error on the spike times
increases exponentially for small input voltages $u_\text{in}$.
\cref{fig:spike_curves} shows that thermal noise typically anticipates the spike time.

The overall measured spike time is thus obtained by combining
the the quantization and the thermal noise time shifts
\begin{equation}
    \label{eq:spike_time_noise}
    \hat{t}_\text{s} = t_s + \delta t_\text{q} - \delta t_\text{thermal} \,,
\end{equation}
that leads to the decoding error
\begin{align}
    \label{eq:u_error}
    \varepsilon(u) &= u - f^{-1}(\hat{t}_\text{s}) \\
    &= u - \frac{u_\text{th}}{1- (1-\frac{u_\text{th}\mp \delta u_\text{thermal}}{u_\text{in} })e^{- \delta t_\text{q}/\tau}}  \,.
\end{align}

\begin{figure*}[t]
    \centering
    \includegraphics[width=\textwidth]{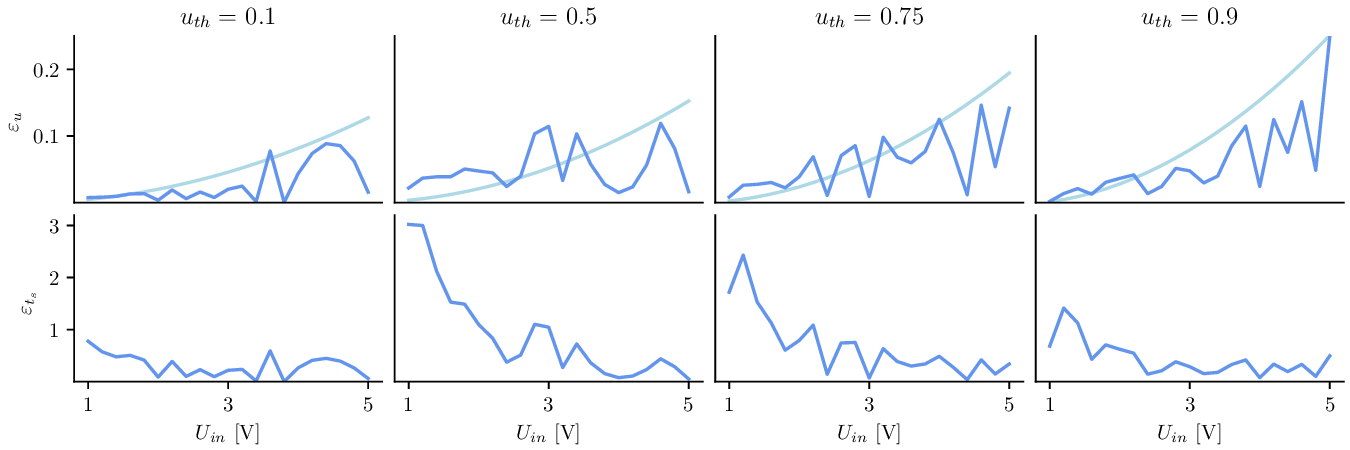}
    \caption{
    Measurement errors of the ASE for encoding constant voltages $u_\text{in} \in \left[1, 5\right]$~V
    with varying $u_\text{th} \in \left[ 0.1, 0.5, 0.75, 0.9 \right]$~V.
    Top: In dark blue, decoding error $\varepsilon_u = |u_\text{in} - f^{-1}(\hat{t_s})|$ for measured spike times $\hat{t_s}$. In light blue, modelled quantization error $\varepsilon_u = |u_\text{in} - f^{-1}(t_s +\delta t_q)|$ with ideal spike time $t_s = f(u_\text{in})$.
    Bottom: normalized error ${\varepsilon_t}_s = |t_s - \hat{t}_s|/T_N$ between the measured spike times $\hat{t_s}$.
    }
    \label{fig:error_plot}
\end{figure*}

\begin{figure}
     \centering
     \begin{subfigure}[b]{0.6\columnwidth}
         \centering
         \includegraphics[width=\textwidth ]{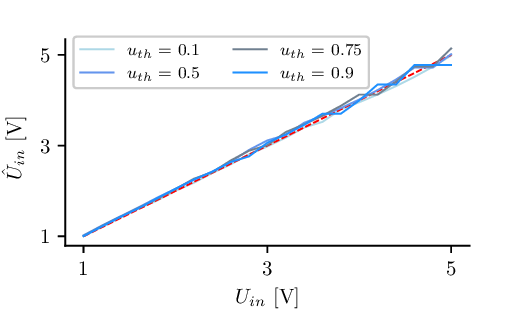}
         \caption{}
         \label{fig:decoded_voltages_ideal}
     \end{subfigure}
     \hfill
     \begin{subfigure}[b]{0.6\columnwidth}
         \centering
         \includegraphics[width=\textwidth ]{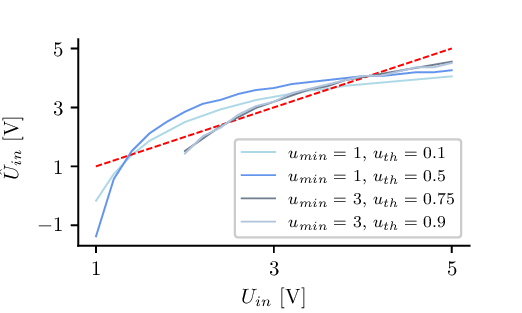}
         \caption{}
         \label{fig:decoded_voltages_linear}
     \end{subfigure}
        \caption{Decoding of the spikes obtained from the ASE when feeding constant input voltages between $1$ and $5$ V,
        when using (a) an ideal decoding scheme for $u_\text{th}$ values of $0.1, 0.5, 0.75,$ and $0.9$ V, respectively;
        and (b) a linear decoding function for $u_\text{th}$ values of $0.1$ and $0.75$ V, respectively.
        Moreover, the second experiment was also run for a narrow voltage input range, namely $2$ to $5$ V.
        The input voltage is shown in red as reference, as it is the target value for the decoding.}
        \label{fig:decoded_voltages}
\end{figure}

\begin{figure*}[h]
    \centering
    \includegraphics[width=\textwidth]{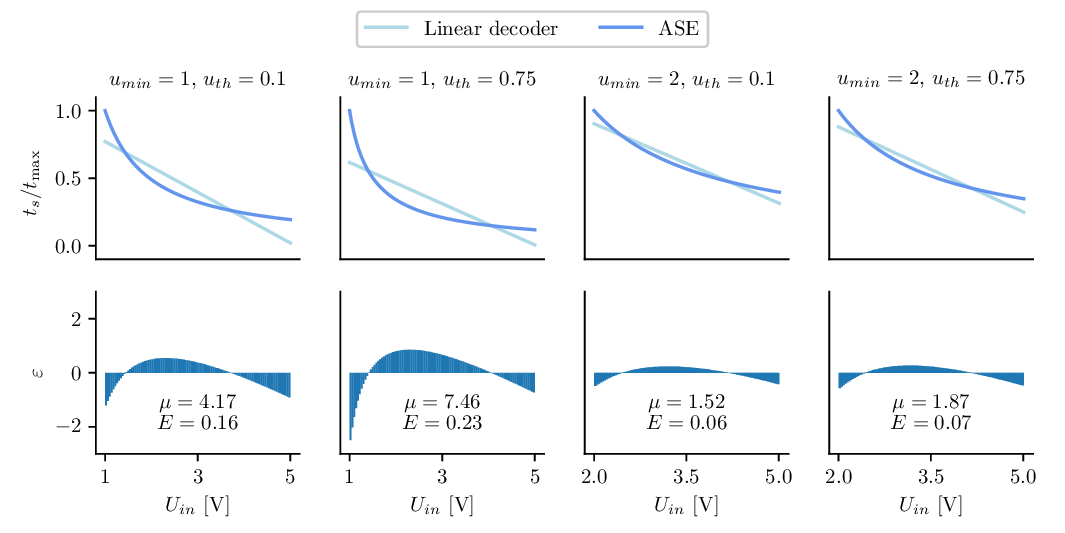}
    \caption{On top, mapping from voltages $U_{in}$ to spike times using the ASE model \eqref{eq:spike_time} and 
    a linear time latency decoder \eqref{eq:ttfs_encoding}, in dark blue and light blue, respectively.
    For ease of comparison, the spike times have been normalized for all experiments, $t_s / t_\text{max}$.
    At the bottom, the decoding error $\varepsilon$ when using \eqref{eq:ttfs_decoding} for decoding the ASE spike times.
    The parameters $t_\text{min}$ and $t_\text{max}$ for the linear decoded are calculated with the optimizer described in section \ref{sec:tuning}
    for minimizing the cumulative decoding error $E$.
    The figure depicts the mapping for $u_\text{min}=1 \, V$ and $2 \, V$, on the two leftmost and two rightmost plots, respectively.
    For each $u_\text{min}$, we tested the ASE for $u_\text{th}=0.1 \, V$ and $0.75 \, V$.}
    \label{fig:sequence_fitting}
\end{figure*}

This experiment converted constant voltages in the working range $[1, 5]$ V to spikes.
We ran the experiment for four different setups with different threshold voltage: \mbox{$u_\text{th} = [0.1, 0.5, 0.75, 0.9]$ V}
(see \cref{fig:error_plot}).

We measured the absolute decoding error as the difference between the input voltage $u_\text{in}$
and the voltage $\hat{u}_\text{in}$ that results from decoding the spike times,
\begin{equation}
    \label{eq:err_voltage}
    \varepsilon_u = | u_{in} - \hat{u}_{in} | \,.
\end{equation}
Moreover, we also measured the error in the spike times 
by comparing the obtained spike times $\hat{t}_s$ and the ideal spike times
$t_s = f(u_\text{in})$ calculated with \eqref{eq:spike_time},
\begin{equation}
    \label{eq:err_time}
    \varepsilon_{t_s} = | t_s - \hat{t}_s | / T_N \,,
\end{equation}
where we use $T_N$ as a normalization coefficient for making the error comparable across experiments.
\cref{fig:error_plot} depicts $\varepsilon_u$ and $\varepsilon_{t_s}$ for the ASE when using the values of $u_\text{th}$ from the previous experiment.
We can see that the quantization error is high for high input voltages,
whereas the thermal noise is low.
On the other hand, low input voltages increase the probability of earlier spikes due to thermal noise,
but low quantization errors in this area make spike times more precise.
These errors are inherent to the used hardware and limit the accuracy of the encoding
regardless of the approach followed for decoding the spikes.

We offer in \cref{fig:decoded_voltages_ideal} a visual comparison between the decoded voltages $\hat{u}_\text{in}$
and the input voltages $u_\text{in}$ for the described experiment.

\subsection{Decoding of flat voltages with  a linear decoder}
\label{sec:exp_linear_decoder}

We tested the ASE setup from the previous section with a linear decoder \eqref{eq:ttfs_decoding}.
We used the optimization strategy introduced in section \ref{sec:tuning}
to find the parameters from \eqref{eq:ttfs_decoding} that best fit the curve from the ASE.
We used a differential evolution method\footnote{\url{https://docs.scipy.org/doc/scipy/reference/generated/scipy.optimize.differential_evolution.html}}
that minimizes $\mathcal{L}$ \eqref{eq:loss}
by adjusting $k_1$ and $k_2$ parameters that fix the decoding time limits
\begin{equation}
    t_\text{lin, min} = t_\text{min}  (1+k_1) \,,
\end{equation}
and
\begin{equation}
    t_\text{lin, max} = t_\text{max}  (1+k_2) \,.
\end{equation}
As the spike times can only be real-valued, the hard boundaries for $k_1$ and $k_2$ are $[-1, \infty]$.
In our implementation, we chose the boundaries $ k_1, k_2 \in [-1, 2]$.

For testing the decoding, we reconstructed the original signal $\hat{u}_\text{in} = g^{-1}(\hat{t}_s)$ from the obtained spike times $\hat{t}_s$
with the inverse of the linear encoding function $g^{-1}$ \eqref{eq:ttfs_decoding}.
We evaluated the reconstruction error in terms of the root mean squared error
between the original signal $u_\text{in}(t)$ and the reconstructed signal $\hat{u}_\text{in}(t)$
\begin{equation}
    \label{eq:mse}
    E = \sqrt{\frac{ \sum_m^M (u_{\text{in}, m} - \hat{u}_{\text{in},m})^2}{M}} \,,
\end{equation}
where $M$ is the number of samples collected during the experiment.

\cref{fig:sequence_fitting} depicts the decoding results for four scenarios with different $u_\text{min}$ and $u_\text{th}$.
A higher $u_\text{min}$ leads to a narrower input range, which directly affects the output dynamic range of the encoder,
i.e., narrow voltage ranges increase the waiting time $t_\text{wait}$ until a spike can take place.
On the other hand, the mapping of the ASE has a smaller curvature, which diminishes the error when using a linear decoder \eqref{eq:ttfs_decoding}.
These two effects are counterbalanced with the threshold voltage, as a higher $u_\text{th}$ leads to a higher dynamic range,
and lower values reduce the decoding error.

After validating the optimizer on the theoretical scenarios,
we applied the fitted curves to data from the experiments with real data in section \ref{sec:ideal_decoder}.
\cref{fig:decoded_voltages_linear} shows the obtained decoded voltages to four scenarios,
with $u_\text{th}=[0.1, 0.75]$ V, and $u_\text{min} = [1, 2]$ V.
Results show that narrowing down the input voltage range is the best way of improving the decoding error
at the expense of reducing the output  dynamic range.
The results for $u_\text{min}=1$ V also show that a higher $u_\text{th}$ leads to a larger curvature
and thus increases the output dynamic range.
This effect is reduced for the results for $u_\text{min}=2$ V because the change in the ratio $u_\text{th} / u_\text{min}$
is smaller for the shown experiments.
Increasing $u_\text{th}$ to values closer to  $u_\text{min}$ would bring more significant improvements to the output dynamic range.

\subsection{Computing the S-FT with the ASE output}
\label{sec:sft_experiment}

We tested the performance of the ASE for an \mbox{end-to-end} pipeline
that computes the frequency spectrum of the input signal.
For the input, we used a function generator that creates signals consisting of one sinusoidal component. These signals can be defined as
\begin{equation}
    \label{eq:sine_function}
    h(t) = A \sin{2 \pi \nu t} + B \,,
\end{equation}
where $A$, $\nu$, and $B$ are the amplitude, frequency, and offset of the sinusoidal wave, respectively.
We connected the function generator to the ASE, generating one spike per sampling period $T_S$.
Finally, we ran the S-FT algorithm on a \mbox{SpiNNaker 2} board
with the spikes generated from the ASE.

We tested the implementation for frequencies in the range $\nu \in [25 \,\text{Hz}, 1 \, \text{kHz}]$.
Same as for the experiment in \cref{sec:exp_linear_decoder}, we created experiments with two different setups for the input voltage range.
Namely, a working range of \mbox{$[1, 5]$ V}, and a working range of \mbox{$[2, 5]$ V},
i.e., we used the parameters $A=2$ and $B=3$, and $A=1.5$ and $B=3.5$, respectively.
We fixed $u_{th}=0.1$ V and $\tau=3$ ms, and the resolution per spike to $N = 100$ time steps.
These parameters lead to maximum spiking times of $315$ and $155 \, \mu$s
for the wide and narrow input voltage ranges, respectively.

\begin{figure*}[t]
    \centering
    \includegraphics[width=\textwidth]{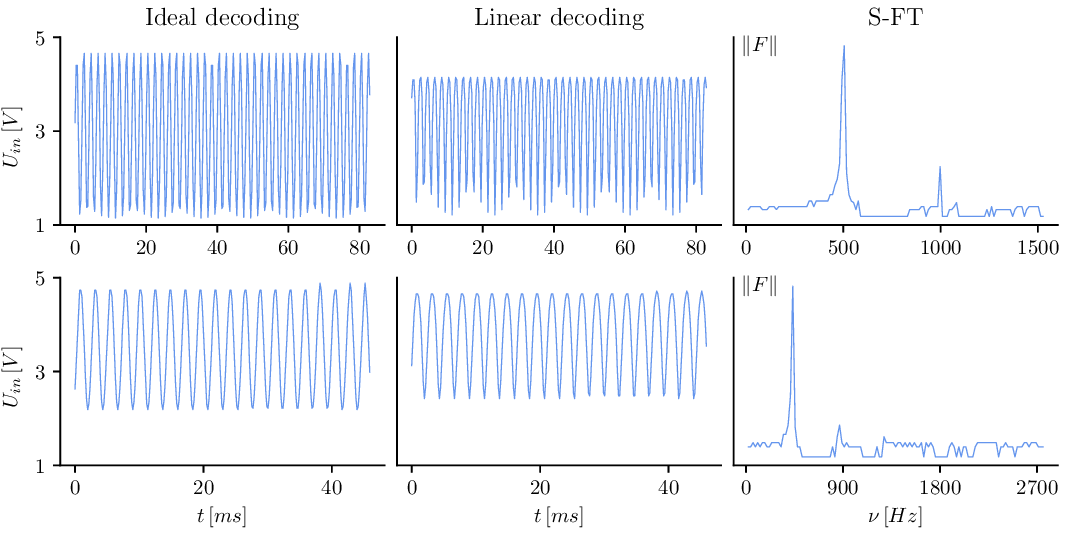}
    \caption{Output of the ASE for a sinusoidal wave of $500 \,Hz$.
    The top plots correspond to a wave with amplitudes ranging between $1 \, \mbox{and} \, 5$ V,
    and the bottom plots correspond to a wave with amplitudes ranging between $2$ and $4$ V.
    From left to right, the plots correspond to the ideal decoding of the spikes, the decoding using a linear function,
    and the result of applying the S-FT on the SpiNNaker 2 board to the output spikes, respectively.
    The ASE was tuned with $\tau=3$ ms and $u_\text{th}=0.1$ V.
    The sampling frequency was $3$ and $5.5$ kHz, respectively, and each sample had a resolution of $100$ time steps.}
    \label{fig:encoded_sinewaves}
\end{figure*}

\begin{figure}[h]
    \centering
    \includegraphics[width=0.6\columnwidth]{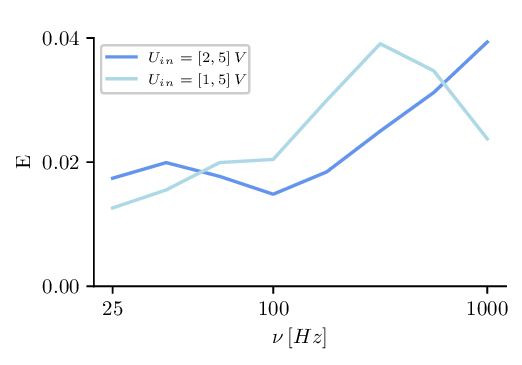}
    \caption{Error of the frequency spectrum when encoding voltages with the ASE and feeding the spikes
    to the S-FT in the SpiNNaker 2 board. The plots show the error for frequencies ranging from $25$ Hz to
    $1000$ Hz, both for wide and narrow input voltage ranges, $V=[1,5]$ V and $[2, 5]$ V, respectively.}
    \label{fig:sft_error}
\end{figure}

In \cref{fig:encoded_sinewaves}, we compare
the result of decoding  the spikes of the ASE using an ideal decoder
and a linear decoder for a wide and narrow input voltage range, respectively.
We also show on the right the result of the S-FT for both input ranges.
The results correspond to a setup with a frequency of $\nu = 500$ Hz.
We repeated this experiment for frequencies \mbox{$ \nu \in [25, 50, 75, 100, 250, 500, 750, 1000]$ Hz}
and calculated the root mean squared error \eqref{eq:mse} between the obtained frequency spectrum
and an FFT computed with a synthetic sine wave sampled with an ideal ADC.
We show the error for the different frequencies in \cref{fig:sft_error}.

%% file: 04-discusion.tex
\section{Discussion}

We modelled the expected output of the proposed ASE from a theoretical perspective 
and validated its behaviour with experiments on analog signals.
In \cref{sec:ideal_decoder} we modelled the error in the spike encoding
as a combination of a quantization error and the thermal error present in analog circuits.
We proved that the quantization error grows exponentially with the value of the input voltage,
and the experiments show evidence that this is the  dominating error  in the output spikes.
The experiment in \cref{sec:ideal_decoder} tested this hypothesis by converting input voltage signals in the range $[1, 5]$ V
to spikes and decoding them back into voltages.
We show the measured error in the decoding on the top row of \cref{fig:error_plot}
and compare it with the quantization error equivalent to half of a time step in the spike times,
i.e., the difference between the input voltage and the voltage that the corresponding spike actually represents.
We also show the errors in the spike times on the bottom row of \cref{fig:error_plot}.
The results indicate that errors in the spike times, mostly due to thermal error,
are larger for low input voltages.
However, these errors do not have a big effect on the decoding error,
as it is mostly dominated by the quantization error.
Finally, \cref{fig:decoded_voltages_ideal} shows the relationship between input and decoded voltages
for the considered voltage range and for four different values of $u_\text{th}$.
We observe that the decoding stays close to the target value, and that errors are more significant towards
high values of $u_\text{in}$.

The experiment in in \cref{sec:exp_linear_decoder} fit a linear decoder to the mapping curve of different setups of the ASE.
This step is necessary before implementing the ASE together with the S-FT,
as the latter assumes phase-encoded input spike times.
We fit a linear decoder for four different ASE setups,
where we vary $u_\text{th}$ and $u_\text{min}$,
and we evaluated the result in terms of the decoding error $E$ and the dynamic output range $\mu$.
We can observe that both parameters have a big influence on both indicators. Large parameter values lead to small errors and  narrow dynamic ranges.
In general, the actual choice of parameters will depend on the requirements and limitations of the application.
If the SNN that processes the spikes is not limited to a linear encoding function
the design should aim to maximize $\mu$, as the decoding error will stay close to zero
(see top and bottom plots in \cref{fig:decoded_voltages}).

The experiment described in \cref{sec:sft_experiment} tested the ASE together with a digital neuromorphic chip
for computing the frequency spectrum of the input analog signal.
As we aimed to minimize the error in the generated spectrum, we chose $u_\text{th} \approx 0.1$ V.
We run experiments for input signal frequencies up to $1$ kHz and $u_\text{min} = 1$ and $2$ V, and fit a linear decoding function for each setup.
\cref{fig:encoded_sinewaves} depicts the results for $\nu=500$ Hz.
The non-linear behaviour of the ASE (\ref{eq:spike_time}) can be approximated by a hyperbolic function
assuming small ratios $u_\text{th}/u_\text{in} \ll 1$, which leads to a compression of the data for low values of $u_\text{in}$,
and an expansion of the encoding range for high values of $u_\text{in}$.
This translates into the appearance of harmonics in the resulting S-FT,
which is specially noticeable for larger input dynamic ranges (see top right plot of \cref{fig:encoded_sinewaves}).
We also plot the total error of the generated frequency spectrum with that of an FFT applied on a synthetic signal
sampled with an ideal ADC, i.e., simulated on a PC without any added noise.
The error stays relatively low, and its mostly due to the usage of a decoding function different to the encoding one.
We observe bigger errors for higher frequencies and higher input voltage ranges.
For $u_\text{in} = [2,  5]$, this trend stops at $\nu = 500$ Hz ,
as after this frequency the second harmonic of the main signal is out of the range of the generated FT.
This source of noise is easy to model, and  this error could be mitigated with a post-processing algorithm.

%% file: 05-conclussion.tex
\section{Conclusion}

We proposed adapting the widely used LIF neuron model for mapping analog signals into spike trains using phase encoding.
By using an adaptive refractory period, the output spike train keeps a notion of the sampling rate,
which is a fundamental requirement for applying FT-based frequency spectrum analysis techniques.

We constructed a circuit prototype and ran it on simple periodic waves to validate the model.
Our experiments show that the proposed approach can encode voltages with high accuracy.
Moreover, we used the output of the ASE for running a spike-based FT on the digital neuromorphic chip \mbox{SpiNNaker 2}.
This implementation is more efficient and simpler than previous approaches,
as it does not require an ADC between the sensor and the digital chip.

Future work shall focus on generating a comprehensive benchmark of the ASE by creating a VLSI version of the circuit
and assessing its performance with complex data obtained from actual sensors.
Those benchmarks shall include critical parameters like energy efficiency and the robustness to the noise of the ASE,
and compare them with the equivalent parameters of ADCs.

%% file: XX-outro.tex
\section*{Acknowledgments}

The authors would like to thank Alexander Lenz and Susanne Junghans for their expertise and input, which was very valuable for implementing and testing the designed circuit.
The authors would also like to thank the Technical University of Dresden for granting them access to the \mbox{SpiNNaker 2} chip and its associated resources.

This  research  has  been  funded  by  the  Federal  Ministry  of Education and Research of Germany in the framework of the KI-ASIC project, with identification number 16ES0995.




\bibliographystyle{ieeetr}
%
\bibliography{main}
%



